\documentclass[proceedings]{JHEP3}

\PrHEP{PrHEP hep2001}
\conference{International Europhysics Conference on HEP}

\usepackage{epsfig}                   

\def\d{{\rm d}}

\def\st#1{\lower.75em\hbox{$|_{#1}$} }
\def\asymp#1{\mathrel{\raisebox{-.4em}{$\widetilde{\scriptstyle #1}$}}}

\newcommand\as{{\alpha_s}}
\newcommand{\amz}{\as(M_{\rm Z^0})}
\newcommand\eps{{\varepsilon}}
\newcommand\LO{leading order }
\newcommand\NLO{next-to-leading order}
\renewcommand\NLO{NLO}
\newcommand\rd{{\rm d}}

\newcommand\rA{{\rm A}}
\newcommand\rB{{\rm B}}
\newcommand\rR{{\rm R}}
\newcommand\rV{{\rm V}}
\newcommand\rLO{{\rm LO}}
\newcommand\rNLO{{\rm NLO}}

\newcommand\beq{\begin{equation}}
\newcommand\eeq{\end{equation}}
\newcommand\beeq{\begin{eqnarray}}
\newcommand\eeeq{\end{eqnarray}}

\newcommand\nlojet{{\tt NLOJET++}}

\title{Progress in QCD next-to-leading order calculations}

\author{\speaker{Zolt\'an Tr\'ocs\'anyi}%
\thanks{Sz\'echenyi fellow of the Hungarian Ministry of Education}\\
University of Debrecen and \\Institute of Nuclear Research of
the Hungarian Academy of Sciences\\ H-4001 Debrecen, PO Box 51, Hungary\\
E-mail: \email{Z.Trocsanyi@atomki.hu}}
\abstract{
I review progress related to the calculation of QCD jet cross sections
at the \NLO\ accuracy. After a short introduction into the theory of
\NLO\ calculations, I discuss two recent developments: the calculation
of two- and three-jet leptoproduction at the \NLO\ accuracy and the
extension of the dipole subtraction method for computing \NLO\
corrections for processes involving massive partons.  
          }

\begin{document}




Jet cross sections at the \NLO\ accuracy consist of the first two terms
of the perturbative expansion in the strong coupling,
\beq
\label{eq:sigma}
\sigma = \sigma^\rLO + \sigma^\rNLO
= \int_m\rd\sigma^\rB + \sigma^\rNLO\:.
\eeq
The \LO\ cross
section is the integral of the fully exclusive Born matrix element of $m$
final-state partons over the available phase space. The \NLO\
correction is a sum of two terms: the real correction is the integral
of the Born matrix element of $m+1$ final-state partons and the virtual
correction is the integral of the interference term between the
Born-level and one-loop amplitudes of $m$ final state partons,
\beq
\label{eq:sigmaNLO}
\sigma^\rNLO = \int_{m+1}\rd\sigma^\rR + \int_m\rd\sigma^\rV\:.
\eeq
In four dimensions both contributions are divergent. An observable is
IR-safe if the total \NLO\ correction is finite in four dimensions.
However, for almost all cases of interest the phase-space integrations
cannot be performed analytically, therefore, in the sum, the
cancellation of the singularities requires special care. In the
literature several general methods are described \cite{QCDatLHC} to
find this finite contribution, each relying on the same principle,
namely we subtract an auxiliary cross section from the real
corrections such that $\d\sigma^\rA$ has the same pontwise singular
behaviour in any dimensions as $\d\sigma^\rR$. Moreover, $\d\sigma^\rA$
has to be chosen simple enough, so that it is analytically integrable
in $d$ dimensions over the one-parton subspaces that cause the soft and
collinear divergences and thus it can be combined with the virtual
contribution to give a finite correction in four dimensions. Thus the
total \NLO\ contribution can be written as a sum of two terms, an
$(m+1)$-parton and an $m$-parton integral, 
\beq
\label{eq:sigmaNLOsub}
\sigma^\rNLO =
\int_{m+1}\!\left[ \left(\rd\sigma^\rR\right)_{\eps=0} -
\left(\rd\sigma^\rA\right)_{\eps=0} \;\right]
+ \int_m\!\left[\rd\sigma^\rV + \int_1 \rd\sigma^\rA \right]_{\eps=0}\:.
\eeq

Although, the principle is simple, the numerical implementation that is
also stable and efficient is difficult. The various methods, that
differ in the explicit choice for the auxiliary cross section, were
developed with the aim of finding the solution with the best numerical
behaviour. In this respect, the dipole method \cite{Catani:1996jh} 
has several advantageous features, only partially shared by the other
techniques:
(i)
The calculation is exact; the same auxiliary cross section is
subtracted and added back.
(ii)
The phase space is exactly the same for the $\d\sigma^\rR$ as for
$\d\sigma^\rA$, which allows for an efficient generation of the phase
space, most suitable for the actual calculation at hand.
(iii)
There is no need for any particular manipulation of the squared matrix
element. 
(iv)
The calculation is Lorentz invariant at any intermediate step, thus
switching between various frames of reference can be achieved by simply
transforming the momenta.
(v)
There is no need for crossing functions, thus partonic cross sections can
be calculated also with hadrons in the initial state.
(vi)
It can be implemented in a fully process independent way, therefore, a truly
general purpose code, suitable for any initial states, can be written and
the same matrix elements in the same code can be used for various processes.
The first such example is the \nlojet\ code that already incorporates
three- and four-jet production in $e^+e^-$ annihilation \cite{Nagy:1997yn},
two- and three-jet leptoproduction \cite{Nagy:2001xb} and two- and three-jet
hadroproduction \cite{Nagy:2001fj}.

As a new application of the method, let us turn to the computation of
\NLO\ corrections to three-jet leptoproduction. Although the \nlojet\
program can be used for calculating any IR-safe observable, for the
sake of comparison with existing data I present differential
distributions that were also measured by the H1 collaboration at HERA. 
In order to make the comparisons, I used the same kinematic range as
employed in the experiment, i.e.,
$5\,{\rm GeV}^2 < Q^2 < 5000\,{\rm GeV}^2$,
$0 < x_{\rm Bj} < 1$,
$0.2 < y < 0.6$,
$-1 < \eta_{\rm jet}^{\rm lab} < 2.5$,
$E_{T,{\rm jet}}^{\rm Breit} > 5\,{\rm GeV}$.
The hard scale was chosen to be the average transverse momentum of the
three-jets. The \LO\ curves were obtained using the CTEQ5L pdf's and
$\as$ run from $\amz = 0.127$ using one-loop running, and the \NLO\
curves with CTEQ5M1 pdf's and $\as$ run from $\amz = 0.118$ using
two-loop running. The three-jet final states were selected using the
inclusive $k_T$ algorithm. The plots in Figs.~\ref{q2dist} and
\ref{xbdist} show the differential distributions in the momentum transfer
squared and the Bjorken variable, respectively. We can observe that the
\LO\ predictions have different shape than the data: too low for small
values of the variables and too high for large values. The radiative
corrections bring theory and experiment much closer. Taking into
account the hadronization corrections, the \NLO\ prediction gives a
remarkably good description of the data. 
\DOUBLEFIGURE
{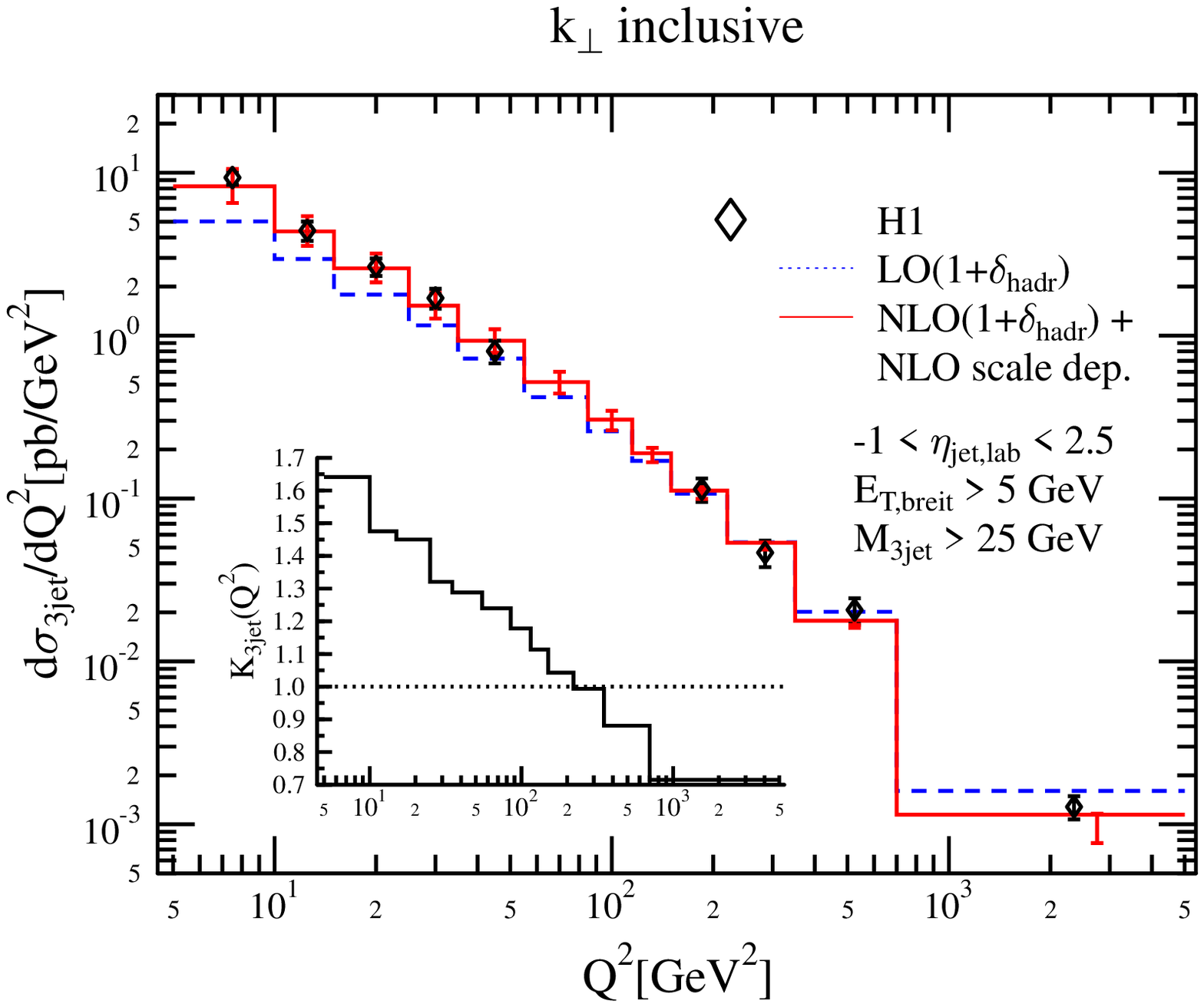,width=.45\textwidth}
{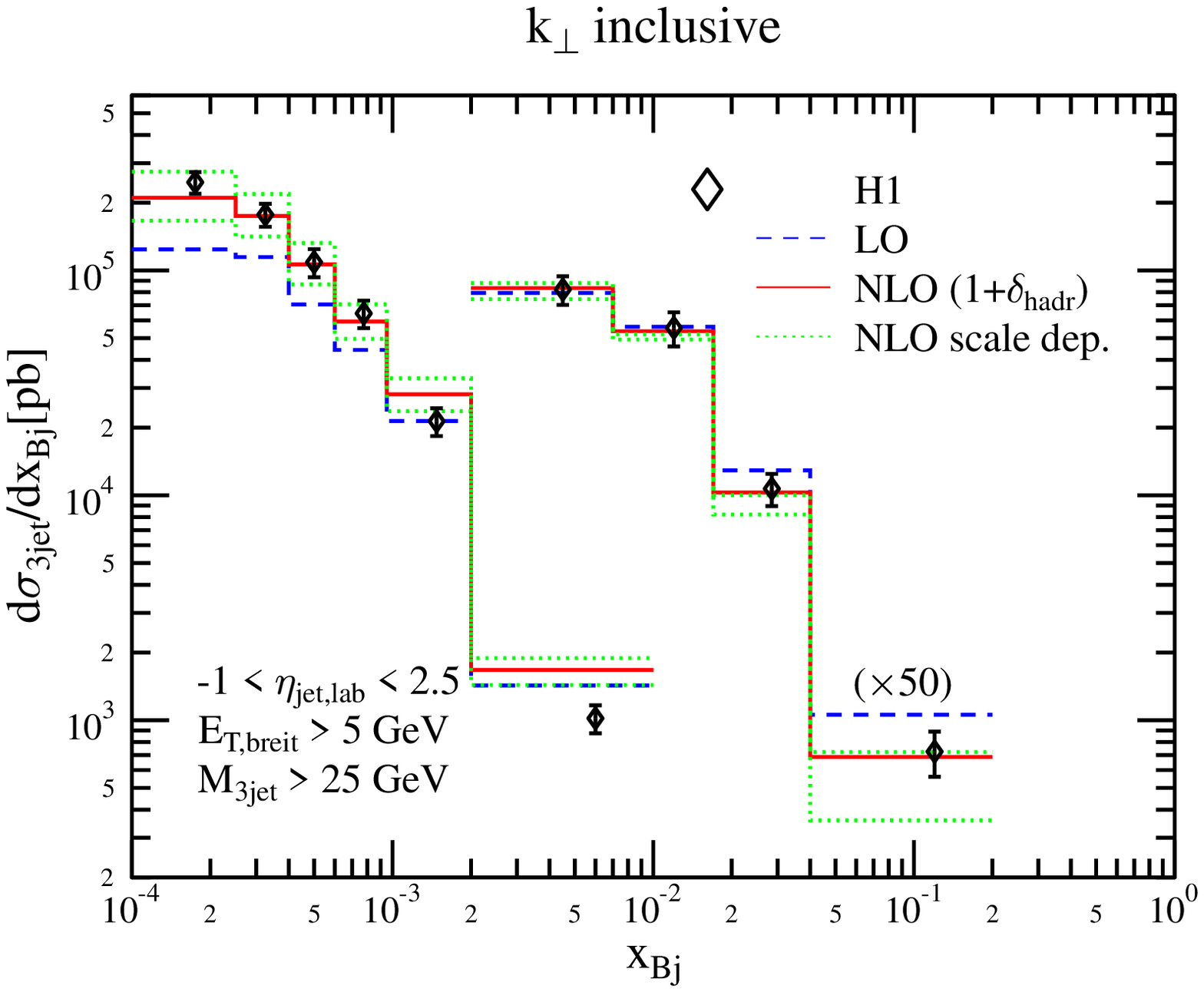,width=.45\textwidth}
{The three-jet differential distribution in the momentum transfer squared
$Q^2$.
\label{q2dist}}
{The three-jet differential distribution in the $x_{\rm Bj}$ Bjorken
variable.\label{xbdist}}

We should consider the impressive agreement with caution. For small
values of $Q^2$ the radiative corrections are rather large (between
30 and 60\,\%), so we may expect that the higher order corrections
are also large, which is also indicated if we look at the dependence of
the \NLO\ prediction for the (3+1)-jet inclusive cross section on the
renormalization and factorization scales in Fig.~\ref{mudeplow}. The scale
dependence of the cross section at \NLO\ is still large indicating
potentially large higher order corrections.  At high $Q^2$ (see
Fig.~\ref{mudephigh} we find that the renormalization-scale dependence
is reduced significantly, while the originally not too strong
factorization-scale dependence does not change substantially. Setting
the two scales equal, at \NLO\ we find a rather flat curve with a wide
plato around the chosen hard scale.  If $Q^2$ is large, however, the
\NLO\ corrections are negative indicating that all order resummation of
terms such as $\ln 1/x_{\rm Bj}$ may be important, which will typically
increase the cross sections.
\DOUBLEFIGURE
{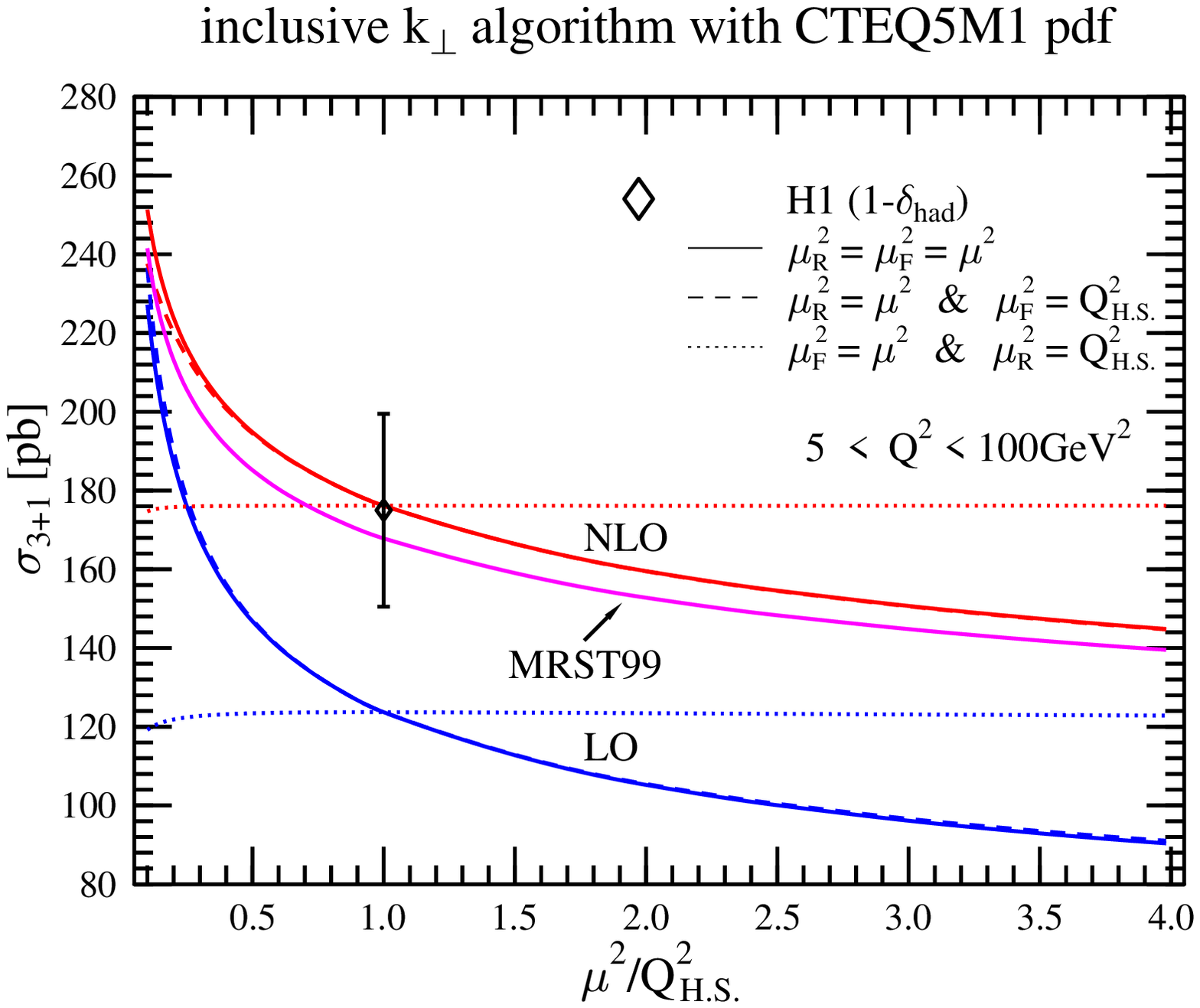,width=.45\textwidth}
{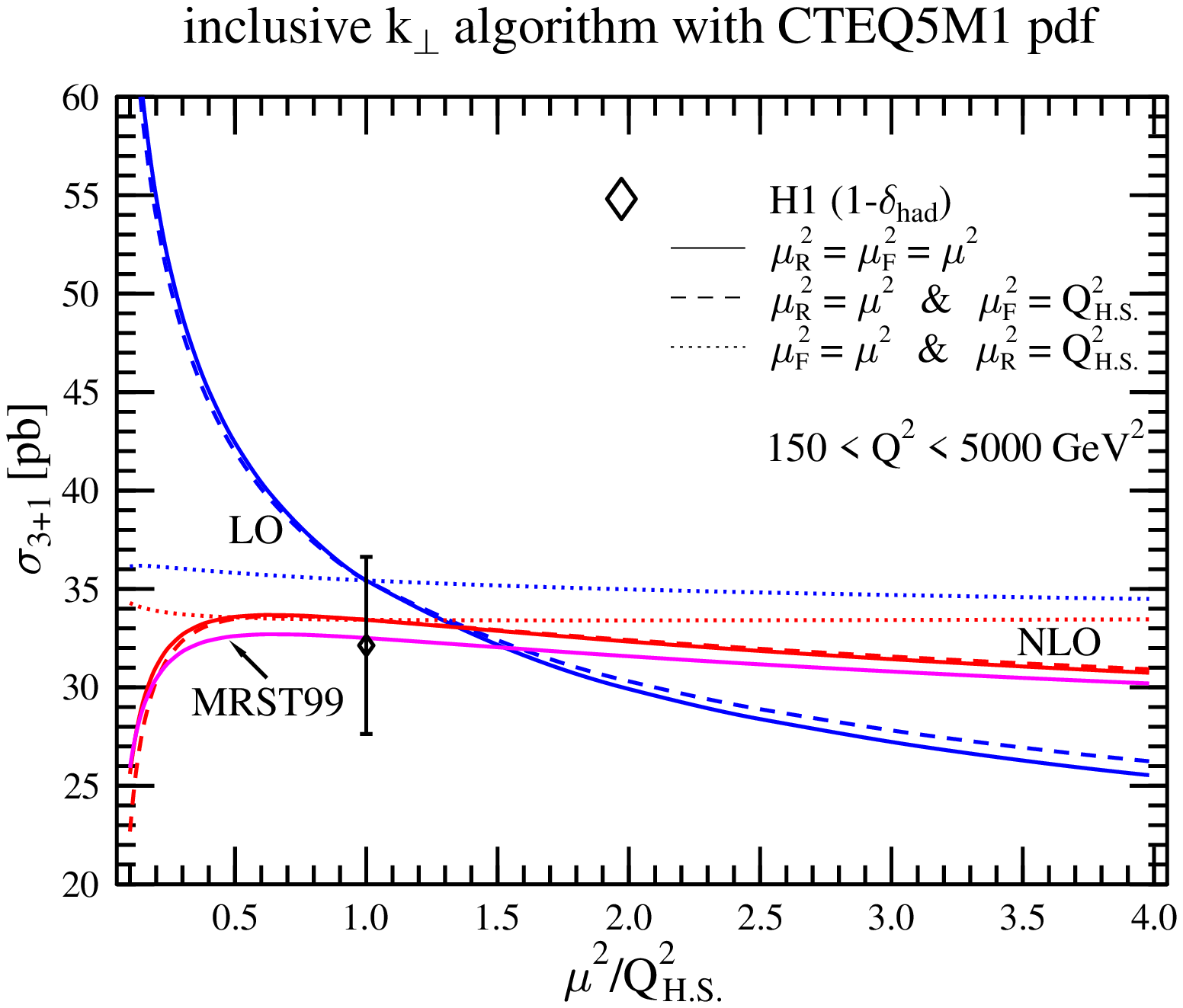,width=.45\textwidth}
{The scale-dependences of the inclusive three-jet cross section at low $Q^2$.
\label{mudeplow}}
{The scale-dependences of the inclusive three-jet cross section at high $Q^2$.
\label{mudephigh}}

Let us now turn to discussing some developments in the theory of \NLO\
calculations for processes involving massive partons.  A lot of such
calculations exist in the literature, but most of them were carried out
for a specific process, not with the general methods worked out for
processes of massless partons. The role of massive partons will
become even more important when the LHC comes into operation and
perhaps a lot of new massive coloured particles will be found, or at
least will be searched for. Recently, both the slicing
\cite{Keller:1999tf} and the dipole \cite{Phaf:2001gc}
techniques were extended to taking into account the parton masses. Here
I advocate another extension of the dipole method which has a smooth
zero-mass limit. 

In order to understand why the zero-mass limit may be important, let us 
recall the different physical roles the finite mass $M$ of the QCD partons
play in different physical processes. In some processes (e.g. the total
cross section for heavy-quark hadroproduction) the finite (and large)
value of $M$ has the essential role of setting the hard scale of the
cross section. In these cases the massless limit is IR {\em unstable},
and the corresponding cross section cannot be computed in QCD
perturbation theory. In other processes (e.g. the production of
heavy-flavoured jets), instead, the hard scale $Q$ is independent of
the mass $M$, and the latter has only the role of an auxiliary (though
important) kinematic scale. These processes are IR {\em stable} in
the massless limit, that is, when $M \to 0$ the cross section is still
infrared- and collinear-safe and, thus, perturbatively computable.

The processes that are perturbatively stable in the massless limit are
often studied in kinematic regions where the typical hard scale $Q$
is much larger than the mass $M$ of one (or more) of the heavy partons.
In this regime the integral of the real term $\rd\sigma^\rR(M)$ of the
NLO cross section in eq.~(\ref{eq:sigmaNLO}) leads to contributions of
the type
\beeq
\label{eq:log}
&&
\int_{m+1}\rd\sigma^\rR(M) \quad \rightarrow \quad
\int_0^{Q^2} \rd{\bf q}_\perp^2 \; \left({\bf q}_\perp^2\right)^{-\eps}
\frac{1}{{\bf q}_\perp^2 + M^2} \asymp{Q \gg M} \ln \frac{Q^2}{M^2}
+ \rm O(\eps)\:,
\\ &&
\label{eq:const}
\int_{m+1}\rd\sigma^\rR(M) \quad \rightarrow \quad
\int_0^{Q^2} \rd{\bf q}_\perp^2 \; \left({\bf q}_\perp^2\right)^{-\eps}
\frac{M^2}{\left[{\bf q}_\perp^2 + M^2\right]^2 } \asymp{Q \gg M}
M^2 \;\frac{1}{M^2} + \rm O(\eps)\:,
\eeeq
where ${\bf q}_\perp$ generically denotes the typical transverse
momentum of the heavy parton with mass $M$.  Since these contributions
are finite when $\eps \to 0$, naively, they would not require any
special treatments within the subtraction method.  However, this could
lead to serious numerical problems in kinematic regions where  $Q \gg
M$. When computing the NLO cross section, the large $\ln Q^2/M^2$
contribution would appear in the first term (the $(m+1)$-parton
integral) on the right-hand side of Eq.~(\ref{eq:sigmaNLOsub}), and it
would be compensated by an equally large (but with opposite sign)
logarithmic contribution arising from the second term (the $m$-parton
integral). Owing to the presence of several large (although
compensating) contributions, a similar naive procedure would lead to
instabilities in {\em any} numerical implementations of the NLO
calculation.  The numerical instabilities would increase by increasing
the ratio $Q/M$ and, in particular, they would prevent from performing
the massless limit.
The second contribution in Eq.~(\ref{eq:const}) can also lead to
numerical instabilities due to the presence of a linearly divergent (in
the limit $M^2/Q^2 \to 0$) integral, because its variance increases
linearly with $Q^2/M^2$.

These numerical problems can be avoided if we set up our massive-parton
formalism by choosing the auxiliary cross section $\rd\sigma^\rA(M)$ in
such a way that the following property is fulfilled:
\beq
\label{eq:smoothlimit}
\lim_{M \to 0}
\int_{m+1}\!\left[ \left(\rd\sigma^\rR(M)\right)_{\eps=0} \! -
\left(\rd\sigma^\rA(M)\right)_{\eps=0} \,\right] =
\int_{m+1}\!\left[ \left(\rd\sigma^\rR(M=0)\right)_{\eps=0} \! -
\left(\rd\sigma^\rA(M=0)\right)_{\eps=0} \,\right]\:.
\eeq
To avoid the problems related to the large logarithmic
contributions in Eq.~(\ref{eq:log}), it is sufficient to impose that
the integral of the subtracted cross section on the left-hand side of
Eq.~(\ref{eq:smoothlimit}) is finite when $M \to 0$. Equation
(\ref{eq:smoothlimit}) is, instead, a stronger constraint. It implies
that, in the evaluation of the subtracted cross section, the massless
limit (or, more generally, the limit $M/Q \to 0$) commutes with the
$(m+1)$-parton integral.
This guarantees that $[ \rd\sigma^\rR(M) - \rd\sigma^\rA(M) ]$ does not
contain integrands of the type in Eq.~(\ref{eq:const}).

The explicit details of the algorithm will be published soon.
Here, I only remark that it follows from eqs.~(\ref{eq:sigmaNLOsub})
and (\ref{eq:smoothlimit}) that the $m$-parton contribution also
possesses a smooth zero-mass limit,
\beq
\lim_{M \to 0}
\int_m\!\left[\rd\sigma^\rV(M) + \int_1 \rd\sigma^\rA(M) \right]_{\eps=0}
=
\int_m\!\left[\rd\sigma^\rV(M=0) + \int_1 \rd\sigma^\rA(M=0)
\right]_{\eps=0}\:,
\label{eq:commute}
\eeq
provided the analytic integration over the one-parton phase space,
$\int_1$ is performed uniformly in the parton masses. Relation
(\ref{eq:commute}) can be utilized to deduce the universal singular
behaviour, both as $\eps \to 0$ and as $M \to 0$, of the one-loop QCD
amplitudes \cite{Catani:2001ef}.
The first computation employing our techniques is published in 
Ref.~\cite{ttH}).  

In summary, the theory of \NLO\ calculations is well established,
including massive partons. For QCD phenomenology the \NLO\ corrections
are known for the most interesting processes, including two- and
three-jet hadro- and leptoproduction and two-jet photoproduction. 
Jet phenomenology could benefit from (i) a program for three-jet
photoproduction, (ii) programs for heavy-flavoured jet lepto-, hadro-,
photoproduction and (iii) programs for vector-boson(s) and associated
jet(s) hadroproduction. 

I thank my collaborators, Z. Nagy (first part) and S. Catani, S.
Dittmaier and M.H. Seymour (second part) for a pleasant a fruitful
collaboration. This work was supported in part by
the EU Fourth Framework Programme
grant FMRX-CT98-0194 (DG 12 - MIHT) and by
the Hungarian Scientific Research Fund grant OTKA T-025482.


\end{document}